# Thermodynamic Function of Life

K. Michaelian,
Instituto de Física, Universidad Nacional Autónoma de México
Cto. de la Investigación Científica
Cuidad Universitaria,
Mexico D.F., C.P. 04510

karo@fisica.unam.mx
Tel: (525)5622-5165, Fax: (525)5616-1535

**Abstract:**
   Darwinian Theory depicts life as being overwhelmingly consumed by a fight for survival in a hostile environment. However, from a thermodynamic perspective, life is a dynamic, out of equilibrium process, stabilizing and coevolving in concert with its abiotic environment. The living component of the biosphere of greatest mass, the plants and cyanobacteria, are involved in the transpiration of vast amounts of water. Transpiration is part of the global water cycle, and it is this cycle that distinguishes Earth from its apparently life barren neighboring planets, Venus and Mars. The water cycle, including the absorption of sunlight in the biosphere, is by far the greatest entropy producing process occurring on Earth. Life, from this perspective, can therefore be viewed as performing an important thermodynamic function; acting as a dynamic catalyst by aiding process such as the water cycle, hurricanes, and ocean and wind currents to produce entropy. The role of animals in this view is that of unwitting but dedicated servants of the plants and cyanobacteria, helping them to grow and to spread into initially inhospitable areas.

**Keywords:** Life, Water Cycle, Non-Equilibrium Thermodynamics

## Introduction

Darwin suggested that life was at the mercy of the forces of Nature and would necessarily adapt through natural selection to the demands of the external environmental. However, it has since become apparent that life plays a pivotal role in altering its physical environment (Lovelock, 1988) and what once appeared to be biotic evolution in response to abiotic pressure is now seen as coevolution of the biotic together with the abiotic to greater levels of complexity, stability, and entropy production (Ulanowicz and Hannon, 1987). Such an understanding, difficult to reconcile within traditional Darwinian theory, fits perfectly well within the framework of non-equilibrium thermodynamics in which dissipative processes spontaneously arise and coevolve in such a manner so as to increase the entropy production of the system plus its environment (Prigogine, 1972, Ulanowicz and Hannon, 1987, Swenson, 1989, Kleidon and Lorenz, 2005, Michaelian, 2005, Michaelian, 2009a).

Life is found everywhere on Earth. On the surface, the components of greatest biomass are the archea, prokaryote, and eukaryote life based on photosynthesis. In the sea, photosynthetic phytoplankton (archea, diatoms, cyanobacteria, and dinoflagallates) can be found in great density (up to $10^9$/ml at the surface) in the euphotic zone which extends to a depth of 50 meters. Almost all photosynthesis ends at the bottom of the Epipelagic zone at about 200 m. Approaching these depths, special pigments are needed to utilize the only faint blue light that can penetrate. On land, diatoms, cyanobacteria, and plants, which evolved from ocean cyanobacteria some 470 million years ago (Wellman and Gray, 2000;

Raven and Edwards, 2001), cover almost every available area, becoming sparse only where conditions are extremely harsh, particularly where liquid water is scarce. Photosynthesizing cyanobacteria have been found thriving in hotsprings at over 70 °C (Whitton and Potts, 2000) and on mountain glaciers and Antarctic ice (Parker *et al*., 1982) where absorption of solar radiation and its dissipation into heat by organic and lithogenic material produces the vital liquid water, even deep within the ice (Priscu *et al*., 2005).

   The thermodynamic driving force for the process of photosynthesis that sustains surface life derives from the low entropy of sunlight and the second law of thermodynamics. Only twenty seven years after Darwin's publication of the theory of evolution through natural selection, Boltzmann (1886) wrote: "The general struggle for existence of animate beings is therefore not a struggle for raw materials – nor for energy which exists in plenty in any body in the form of heat -- but a struggle for entropy, which becomes available through the transition of energy from the hot sun to the cold earth". In photosynthesis, high-energy photons in the visible region of the Sun's spectrum are converted by the chloroplasts into low energy photons in the infrared region. Part of the free energy made available in the process is utilized to maintain and propagate life. In this manner, photosynthetic life obtains its sustenance through the conversion of the low entropy of sunlight into the higher entropy of heat and thereby contributes to the positive entropy production of the Earth as a whole.

 However, the proportion of the Sun's light spectrum utilized in photosynthesis is small and thus the entropy producing potential of photosynthesis is small. Gates (1980) has estimated that the percentage of available (free) energy in solar radiation that shows up in the net primary production of the biosphere is less than 0.1%. Respiration consumes a similarly small quantity (Gates, 1980). Of all the irreversible processes performed by living organisms, the process generating by far the greatest amount of entropy (consuming the greatest amount of free energy) is the absorption of sunlight by organic molecules in the presence of water leading to *evapotranspiration*. Great quantities of water are absorbed by the root systems of plants and brought upwards to the leaves and then evaporated into the atmosphere. More than 90% of the free energy available in the sunlight captured by the leaves of plants is used in transpiration. In the oceans, phytoplankton within the euphotic zone absorb sunlight and transform it into heat that can be efficiently absorbed by the water. The temperature of the ocean surface is thereby raised by phytoplankton (Kahru et al., 1993) leading to increased evaporation, thereby promoting the water cycle.

There appears to be no inportant physiological need for the vast amount of transpiration carried out by land plants. It is known that only 3% of the water transpired by plants is used in photosynthesis and metabolism. In fact, most plants can grow normally under laboratory conditions of 100% humidity, at which the vapor pressure in the stoma of the leaves must be less than or equal to that of the atmosphere, and therefore transpiration is necessarily zero (Hernández Candia, 2009). Transpiration has often been considered as an unfortunate by-product of the process of photosynthesis in which water is unavoidably given off through the stoma of plants which are open in order to exchange $CO_2$ and $O_2$ with the atmosphere (Gates, 1980). Plants consist of up to 90% water by mass and thus appear to expose themselves to great risk of drying by transpiring so much water. Others have argued that transpiration is useful to plants in that it helps to cool its leaves to a temperature optimal for photosynthesis. Such an explanation, however, is not convincing since Nature has produced examples of efficient photosynthesis at temperatures of up to 70 °C (Whitton and Potts, 2000). In any case, there exists other simpler and less free energy demanding strategies to reduce leaf temperature such as smaller or less photo-absorbent leaves. On the contrary, the evolutionary record indicates that plants and phytoplankton have evolved new pigments to absorb ever more completely the Sun's spectrum. Dense pine forests appear black in the midday sun. Most plants appear green, not so much for lack of absorption at these wavelengths, as for the fact that the spectral response of human eyes peaks precisely at these wavelengths (Chang, 2000).

Transpiration is in fact extremely free energy intensive and, according to Darwinian Theory, such a process, with little direct utility to the plant, should have been eliminated or suppressed through natural selection. Plants which are able to take in $CO_2$ while reducing water loss, by either opening their stoma only at night (CAM photosynthesis), or by reducing photorespiration ($C_4$ photosynthesis, see below), indeed have evolved 32 and 9 million years ago respectively (Osborne and Freckleton, 2009). However, the water conserving photosynthesis has not displaced the older, heavily transpiring $C_3$ photosynthesis which is still relevant for 95% of the biomass of Earth. Instead, new ecological niches in water scarce areas have opened up for the CAM and $C_4$ plants, as, for example, the cacti of deserts.

All irreversible processes, including living systems, arise and persist to produce entropy. This is not incidental, but rather a fundamental principle of Nature. Excessive transpiration has not been eliminated from plants, despite the extraordinary free energy costs, precisely because the basic thermodynamic function of a plant is to increase the global entropy production of the Earth and this is achieved by dissipating high energy photons in the presence of water and thereby augmenting the global water cycle.

**The Water Cycle**
Absorption of sunlight in the leaves of plants may increase their temperature by as much as 20°C over that of the ambient air (Gates, 1980). This leads to an increase of the $H_2O$ vapor pressure inside the cavities of the leaf with respect to that of the colder surrounding air. $H_2O$ vapor diffuses across this gradient of chemical potential from the wet mesophyll cell walls (containing the chloroplasts), through the intercellular cavities, and finally through the stoma and into the external atmosphere. There is also a parallel, but less efficient, circuit for diffusion of $H_2O$ vapor in leaves through the cuticle, providing up to 10% more transpiration (Gates, 1980). The $H_2O$ chemical potential of the air at the leaf surface itself depends on the ambient relative humidity and temperature, and thus on such factors as the local wind speed and insolation. Diffusion of $H_2O$ vapor into the atmosphere causes a drop in the water potential inside the leaf which provides the force to draw up new water from the root system of the plants.

Evaporation from moist turf (dense cut grass) can reach 80% of that of a natural water surface such as a lake (Gates, 1980), while that of a tropical forest can often surpass by 200% that of such a water surface (Michaelian, 2009b). Single trees in the Amazon rain forest have been measured to evaporate as much as 1180 liters/day (Wullschleger *et al.*, 1998). This is principally due to the much larger surface area for evaporation that a tree offers with all of its leaves. Natural water surfaces, in turn, evaporate approximately 130% of distilled water surfaces due to the increased UV and visible photon absorption at the surface as a result of phytoplankton and other suspended organic materials, including a large component (up to $10^9$/ml at the surface) of viral and dissolved DNA resulting from viral lysing of bacteria (Wommack and Colwell, 2000).

The water vapor transpired by the leaves, or evaporated by the phytoplankton, rises in the atmosphere, because water vapor at 0.804 g/l is less dense than dry air at 1.27 g/l, to a height corresponding to a temperature of about 259 K (-14 °C) (Newell *et al.*, 1974) at which it condenses around suspended microscopic particles forming clouds. Over oceans, an important constituent of these microscopic particles acting as seeds of condensation are the sulfate aerosols produced by the oxidation of dimethylsulfide released by the phytoplankton themselves (Charlson *et al.*, 1987). Condensation of the water releases an amount of latent heat of condensation ($2.427 \times 10^6$ J/kg) into the upper atmosphere, much of which is then radiated into outer space at infrared wavelengths. In this manner, the Earth maintains its energy balance with space; the total energy incident on the biosphere in the form of sunlight is approximately equal to the total energy radiated by the biosphere into space at infrared wavelengths. Energy is conserved while the entropy of the Universe is augmented in the process.

The formation of clouds may at first consideration seem to have a detrimental effect on the water cycle since cloud cover on Earth reflects approximately 20% of light in the visible region of the Sun's spectrum (Pidwirny and Budicova, 2008), thereby reducing the potential for evaporation. However, evapotranspiration is a strong function of the local relative humidity of the air around the leaves of plants or above the surface of the oceans. By producing regions of local cooling during the day on the Earth's surface, clouds are able to maintain the average wind speed at the Earth's surface within dense vegetation (see for example, Speck (2003)) at values above the threshold of 0.25 m/s required to make the boundary-layer resistance to water loss almost negligible in a plant leaf, thus procuring maximal transpiration (Gates, 1980).

Sublimation and ablation of ice over the polar regions, promoted in part by photon absorption of cyanobacteria within the ice, is also important to the water cycle, evaporating up to 30 cm of ice per year (Priscu *et al.*, 2005).

**Production of Entropy**
The driving force of all irreversible processes, including the water cycle, is the *production of entropy*. The basic entropy producing process occurring on Earth is the absorption and dissipation of high energy photons to low energy photons, facilitated in part by the plants and cyanobacteria in the presence of water. The global entropy production of the Earth can be determined by considering the change in the frequency $\nu$ distributions of the radiation incident from the Sun, $I_{incident}(\nu)$, and that radiated by the Earth, $I_{radiated}(\nu)$ (Ulanowicz and Hannon, 1987). The flow of photons can be considered as an ideal gas of Bose-Einstein particles for which the internal temperature $T$ can be related to the frequency $\nu$ by the relation $kT = h\nu$, where $k$ is Boltzmann's constant and $h$ is Planck's constant. Gibb's equation for the flow of entropy $dS$ during time interval $dt$ at a particular frequency $\nu$ is (Callen, 1985),

$$\frac{dS(\nu)}{dt} = \frac{1}{T(\nu)} \frac{dE(\nu)}{dt} = \frac{I(\nu)}{T(\nu)} \tag{1}$$

where $I(\nu) \equiv \frac{dE(\nu)}{dt}$ is defined as the energy flow or irradiance at frequency $\nu$, and where we have neglected the volume and chemical potential terms in the Gibb's equation since the volume of the entire Earth-space system is constant and the chemical potential for photons is zero (Callen, 1985). The global production of entropy of the Earth in its interaction with its solar environment is then just the difference of the radiated to incident entropy flow integrated over all frequencies

$$P = \int_0^\infty \frac{dS_{radiated}(\nu)}{dt} - \frac{dS_{incident}(\nu)}{dt} d\nu .$$

Using equation (1) with $T(\nu) = h\nu/k$ gives

$$P = \frac{k}{h} \int_0^\infty (I_{radiated}(\nu) - I_{incident}(\nu))/\nu \, d\nu . \tag{2}$$

A very approximate measure of this entropy production for the entire Earth can be obtained by making a black body assumption for the incident and radiated irradiances (Aoki, 1983). The Planck distribution law for the radiation emitted per unit area, per unit solid angle, per unit frequency, for a blackbody at temperature $T$ is (Landau and Lifshitz, 1988)

$$I(\nu)\big|_T = \frac{2h\nu^3}{c^2} \frac{1}{e^{h\nu/kT} - 1} . \tag{3}$$

Taking, $T_{raduated}$ as the average temperature of the Earth surface, 287 K (14°C), and $T_{incident}$ as the temperature of the surface of the Sun, 6073 K (5800 °C), Eq. (2) with Eq. (3) gives as an approximation for the net entropy production of the Earth of $1.19 \times 10^{-4}$ J cm$^{-2}$ s$^{-1}$ K$^{-1}$, about 50% greater than that of Earth's neighboring apparently lifeless planets of Venus and Mars (Aoki, 1983).

Equation (2) with Eq. (3) demonstrates that there is greater potential for entropy production for absorption of high frequency light in the biosphere than for the absorption of low frequency light. It is thus probably not coincidental that the present day atmosphere of the Earth has evolved to one of relatively low albedo and high atmospheric transparency such that the highest frequencies (and most intense) part of the Sun's spectrum can arrive at the biosphere and be efficiently dissipated by organic molecules in contact with water. It has been suggested, for example, that there probably existed a thick organic haze, of high albedo in the visible, during part of the Archean from 2.9 – 2.7 Ga., responsible for global glaciations during this period (Zahnle et al., 2007). Organic haze is preferentially produced through ultraviolet photoreactions on an atmosphere with a methane over carbon dioxide ratio greater than or equal to one ($CH_4/CO_2 \geq 1$) which probably occurred ca. 2.9 Ga. (Lowe and Tice, 2004, Zahnle et al., 2007). Such a large ratio could have arisen due to enhanced weathering associated with continent formation and the production of calcium carbonates, causing depletion of atmospheric carbon dioxide. The organic haze existing today on Titan, for example, permits less than 10% of the visible light to reach the surface (Coustenis and Taylor, 1999). With the spread of oxygen producing photosynthetic microbes in the late Archean ca. 2.5 Ga., however, the $CH_4/CO_2$ ratio was driven below the critical value for haze formation due to oxidation of $CH_4$. Such a scenario, suggesting the importance of life in reducing the albedo of Earth's early atmosphere, can also explain two well established facts from the era; the green house effect necessary to explain the evidence for liquid water given a faint young Sun (the faint young Sun paradox), and evidence for glaciations around 2.9 Ga. (Lowe and Tice, 2004).

Comparison of the approximate calculations of Aoki (1983) for the entropy production of the different planets suggests that the biosphere may play a particularly important role in the entropy production of Earth. About 51% of the energy arriving from the Sun in short wave radiation is absorbed in the biosphere (at the surface of the Earth), the rest being absorbed by the clouds and upper atmosphere (19%), reflected by the clouds or surface (24%), or scattered by the atmosphere back into space (6%) (Pidwirny and Budicova, 2008). About half of the available energy arriving at the Earth's surface is used to evaporate the great quantities of water that is eventually returned to the Earth's surface in the form of rain. The other half is roughly equally divided between driving ocean and wind currents. As already mentioned, a negligible proportion of the free energy absorbed by the biosphere (0.1%) goes into the metabolism and production of biomass. However, most of the visible and near ultraviolet spectra of sunlight (where the Sun is most intense in terms of free energy) is not readily absorbed by pure water, as can be deduced by the transparency of water at these wavelengths (with the exception of a small region in the far ultraviolet absorbed by the electronic excitation of the oxygen-hydrogen covalent bonds). Only infrared light can be efficiently absorbed by water and transferred to the vibrational, and a lesser fraction ($< 10^{-4}$) to the rotational, degrees of freedom of the water molecules. This absorbed vibrational energy can then cause the breaking of hydrogen bonds binding water molecules and thereby facilitate evaporation at the water surface.

Organic molecules, due to the nature of the strong electronic covalent bonding, are efficient absorbers of sunlight in the visible and ultraviolet regions of the Sun's spectrum. The chlorophyll molecule and associated pigments absorb in the visible region between approximately 400 nm and 700 nm, with chlorophyll A peaking in absorption at 410 nm and 680 nm. The nucleic acids and proteins containing amino acids with aromatic rings (Trp, Tyr, Phe) are particularly potent absorbers of ultra violet light within the 200-300 nm region due to the $\pi \to \sigma$, $\pi \to \sigma^*$, and $\sigma \to \sigma^*$ electronic transitions, with peak absorption for the nucleic acids at 260 nm and that for proteins at 280 nm (Chang, 2000).

Mycosporine-like amino acids (MAA's) found in phytoplankton absorb across the UVB and UVA regions (310-400 nm) (Whitehead and Hedges, 2002). The amount of ultraviolet light reaching the Earth's surface today, particularly in the <290 nm UVC wavelength region, is very small compared to that of UVB+UVA and visible light due to absorption by ozone and $O_2$ and thus this region plays a very small part in the entropy production associated with the water cycle, but this may not have been the case at the beginnings of life on Earth (Michaelian, 2009a). During the archean, the Sun was more active in the ultraviolet and the Earth's atmosphere was more reflective and absorptive in the visible while less so in the ultraviolet. This may have been due to a high layer of sulfuric acid clouds as on Venus today, the result of UV photochemical reactions with the most common volcanic gases of $SO_2$, $CO_2$, and $H_2O$, or to clouds of water, or to organic haze as on Titan today, the result of UV photochemical reactions on $CO_2$ and $CH_4$ (Lowe and Tice, 2004).

**The Importance of Life to the Water Cycle**
The very existence of liquid water on Earth can be attributed to the existence of life. Through mechanisms related to the regulation of atmospheric carbon dioxide first espoused in the Gaia hypothesis (Lovelock, 1988), life is able to maintain the temperature of the Earth within the narrow region required for liquid water, even though the amount of radiation from the Sun has increased by about 25% since the beginnings of life (Newman and Rood, 1977, Gough, 1981). Physical mechanisms exist that disassociate water into its hydrogen and oxygen components, for example through photo-dissociation of water by ultraviolet light (Chang, 2000). Photo-dissociation of methane has been suggested as a more important path to loosing the hydrogen necessary for water (Catling et al., 2001). Free hydrogen, being very light, can escape Earth's gravity and drift into space, being dragged along by the solar wind. This loss of hydrogen would have lead to a gradual depletion of the Earth's water (Lovelock, 2005). However, photosynthetic life sequesters oxygen from carbon dioxide thereby providing the potentiality for its recombination with the free hydrogen to produce water. For example, hydrogen sulfide is oxidized by aerobic chemoautotrophic bacteria, giving water as a waste product (Lovelock, 1988). Oxygen released by photosynthetic life also forms ozone in the upper atmosphere which protects water vapor and methane in the lower atmosphere from ultraviolet photo-dissociation. In this manner, the amount of water on Earth has been kept relatively constant since the beginnings of life.

It has been estimated that about 496,000 $km^3$ of water is evaporated yearly, with 425,000 $km^3$ (86%) of this from the ocean surface and the remaining 71,000 $km^3$ (14%) from the land (Hubbart and Pidwirny, 2007). Evaporation rates depend on numerous physical factors such as insolation, absorption properties of air and water, temperature, relative humidity, and local wind speed. Most of these factors are non-linearly coupled. For example, local variations in sea surface temperature due to differential photon absorption rates caused by clouds or local phytoplankton blooms, leads to local wind currents. Global winds are driven by latitude variation of the solar irradiance and absorption, and the rotation of the Earth. Relative humidity is a function of temperature but also a function of the quantity of microscopic particles available for seeds of condensation (a significant amount of which are supplied by biology (Lovelock, 1988)).

The couplings of the different factors affecting the water cycle imply that quantifying the effect of biology on the cycle is difficult. However, simulations using climate models taking into account the important physical factors have been used to estimate the importance of vegetation on land to evapotranspiration. Kleidon (2008) has shown that without plants, average evaporation rates on land would decrease from their actual average values of 2.4 mm/d to 1.4 mm/d, suggesting that plants may be responsible for as much as 42% of the actual evaporation over land.

There appears to be little recognition in the literature of the importance of cyanobacteria and other organic matter floating at the ocean surface to evaporation rates. Irrespective of other factors such as wind speed and humidity, evaporation rates should be at least related to the energy deposited in the sea surface layer. A calculation can therefore be made of the effect of biology on the evaporation rates over oceans and lakes.

Before attempting such a calculation, it is relevant to review the biological nature of the air-sea surface interface, and energy transfer within this layer, based on knowledge that has emerged over the last decade. This skin surface layer of roughly 1 mm thickness has its particular ecosystem of high density in organic material (up to $10^4$ the density in water slightly below (Grammatika and Zimmerman, 2001)). This is due to the scavenging action of rising air bubbles due to breaking waves, surface tension, and natural buoyancy (Grammatika and Zimmerman, 2001). The organic material consists of cyanobacteria, diatoms, viruses, free floating RNA/DNA, and other living and non-living organic material such as chlorophyll and other pigments. Most of the heat exchange between the ocean and atmosphere of today occurs from within this upper 1 mm of ocean water. For example, most of the radiated infrared radiation from the sea comes from the upper 100 $\mu$ m (Schlussel, 1999). About 52% of the heat transfer from this ocean layer to atmosphere is in the form of latent heat (evaporation), radiated longwave radiation accounts for 33%, and sensible heat through direct conduction accounts for the remaining 15%.

During the day, infrared (700-10000 nm), visible (400-700 nm), and ultraviolet (290-400nm) light is absorbed at the sea surface. In the NE Atlantic, for example, daytime temperatures at the skin surface have been measured to increase on average by 2.5K (up to 4.0K) compared to the practically constant temperature at an ocean depth of 10m (Schlüssel et al., 1990). Nighttime temperatures at the surface, on the contrary, are decreased on average by 0.5K (up to 0.8K) with respect to the constant temperature at a depth of 10m. It is thus of interest to determine how much of this day-tine heating is due to the organic material in this layer, and the relative contributions due to UV, visible, or infrared radiation. Such a determination will allow an estimate of the effect of life on the evaporation of water over oceans and entropy production under different physical conditions, for example, under cloudy or clear skies. For the sake of calculation, we take the surface skin layer for light absorption and heat exchange to the atmosphere to be 1mm (this should be an upper limit for the relevant thickness for energy exchange since below this depth turbulence and mixing with lower ocean depths becomes relevant (Soloviev and Lukas, 2006).

Three distinct wavelength regions are considered for the calculation; 290-400 nm (UV) (below 290 nm almost all light is blocked by $O_3$), 400-700 nm (visible), and 700-10000 nm (infrared). The blackbody spectrum of the Earth at 288K peaks at 10000 nm so absorption at greater wavelengths than this would not contribute to net heating (there is, in any case, very little energy in sunlight beyond this wavelength). We first calculate the total amount of energy arriving at the sea surface in each wavelength region for a clear sky with no clouds and the sun directly overhead. This can be obtained by integrating the area under a plot of the irradiance at the Earth's surface as a function of wavenumber, such as that given by Gates (1980, Fig. 8.17). The result is given in the first row of table 1.

To calculate the amount of energy deposited per unit time in each wavelength region within the 1mm skin layer of pure ocean water without organic material, we use an average water absorption coefficient for the middle of the UV (345 nm) and visible (550) wavelength range, whereas for the infrared region we use the absorption value at 1050 nm since this corresponds to the greatest incident contribution not absorbed by water vapor in the atmosphere (Gates, 1980, Fig. 8.17), and because the irradiance drops off sharply at greater wavelengths. Chaplin (2009) give the following absorption coefficients for pure water; $A_0^{345} = 1.0 \times 10^{-4}$ cm$^{-1}$ $A_0^{550} = 8.0 \times 10^{-4}$ cm$^{-1}$ $A_0^{1050} = 0.2$ cm$^{-1}$ (Fig. 4).

The flux of energy deposited in the skin layer is then
$$\Delta I = (I_0 - I(x)) = I_0(1 - \exp(-Ax)) \quad (4)$$
with $x = 0.1$ cm and $A$ is the relevant absorption coefficient. The results are given in the second row of table 1.

To calculate the amount of energy per unit time deposited within the 1mm skin layer of ocean water *with* organic material for each wavelength region, we need the absorption coefficients for the ocean surface microlayer at the different wavelengths. Unfortunately, there do not appear to be any published data in this regard, however, Grammatika and Zimmerman, (2001) suggest that the skin microlayer contains up to $10^4$ times the density of organic material as water slightly below. This factor of $10^4$ is an order of magnitude greater than that of the ratio between the densities of organic matter in very turbid costal waters to that of deep sea water (Wommack and Colwell, 2000). We therefore take the absorption coefficients for costal turbid waters obtained for the Baltic Sea from Bricaud et al. (1981, Fig. 3) as being a lower limit surrogate to that of the surface skin layer of the ocean, giving $A_{org}^{345} = 0.1$ cm$^{-1}$, $A_{org}^{550} = 8.0 \times 10^{-3}$ cm$^{-1}$, $A_{org}^{1050} = 0.2$ cm$^{-1}$
The value for the infrared absorption is the same as that for pure water since organic molecules absorb very little compared to water in this wavelength region. Using equation (4) with these values we obtain the third row of table 1.

By comparing the third row with the second row of table 1 it can be determined that on a clear day, with the sun directly overhead (air mass of 1.0), the organic matter floating in the surface skin layer increases the absorption of energy in this layer by about 13% over what its value would be without this organic matter. Since the rate of evaporation would be proportional to the energy absorbed, it can be estimated that organic material in the sea surface microlayer increases the evaporation from the surface by roughly this same percentage. This is consistent, although somewhat lower, than the increase in the measured latent outgoing heat flux for nutrient induced phytoplankton blooms in an enclosed area of a lake as determined by Jones et al. (2005). The absorption of UV light contributes more than double the amount of that due to absorption of visible light.

The relative contribution to the entropy production in the sea surface microlayer resulting from photon absorption and dissipation in each wavelength region can now be approximated by calculating the increase in the number of energy microstates resulting when a photon of the given wavelength is dissipated into many more photons of wavelength 10,000 nm, corresponding to the wavelength of the peak in the black-body spectrum of the Earth at 288 K (Gates, 1980). The relative number of 10,000 nm photons produced by absorption and dissipation in each wavelength region can be obtained by simply dividing the energy of a photon at the central wavelength in each wavelength region by the energy of a 10,000 nm photon and then multiplying by the flux of energy deposited (row 3 of Table 1) in the microlayer in this wavelength region. The Boltzmann relation gives the entropy produced as just proportional to the natural logarithm of this number of created microstates. Dividing this value in each wavelength region by the sum total for all regions gives the percentage contribution to the total for each region. The result is given in the fourth row of Table 1. By absorbing and dissipating UV and visible light, the organic matter in the sea surface microlayer contributes about 33.2 + 20.4 = 53.6% to the entropy production in this layer (fourth row of table 1).

| CLEAR SKIES | UV (290-400 nm) | Visible (400-700 nm) | Infrared (700-10000 nm) |
|---|---|---|---|
| Energy flux reaching Earth's surface (total 1029.3 W/m$^2$ ) | 50.5 W/m$^2$ direct 33.7 W/m$^2$ skylight -------------- 84.2 W/m$^2$ Global (8.2%) | 428.2 W/m$^2$ direct 53.9 W/m$^2$ skylight -------------- 482.1 W/m$^2$ Global (46.8%) | 456.3 W/m$^2$ direct 6.7 W/m$^2$ skylight ----------- 463.0 W/m$^2$ Global (45%) |
| Energy flux absorbed in 1 mm skin layer (pure water). | 0.84x10$^{-3}$ W/ m$^2$ ($A_0^{345} = 1.0x10^{-4}$ cm$^{-1}$) | 38.6x10$^{-3}$ W/m$^2$ ($A_0^{550} = 8.0x10^{-4}$ cm$^{-1}$) | 9.16 W/m$^2$ ($A_0^{1050} = 0.2$ cm$^{-1}$) |
| Energy flux absorbed in 1 mm skin layer (ocean water with organic material). | 0.84 W/m$^2$ ($A_{org}^{345} = 0.1$ cm$^{-1}$) | 0.39 W/m$^2$ ($A_{org}^{550} = 8.0x10^{-3}$ cm$^{-1}$) | 9.16 W/m$^2$ ($A_{org}^{1050} = 0.2$ cm$^{-1}$) |
| % of total entropy production in skin. | 33.2 | 20.4 | 46.4 |

Table 1. Values of energy deposition per unit time in the sea surface microlayer and the contribution to entropy production for the different wavelength regions assuming clear skies.

In the case of an overcast day, proportionately less infrared radiation arrives at the ocean surface because of the strong absorption of infrared by water drops in clouds (see Fig. 8.17 of Gates, 1980). We take the absorption coefficients corresponding to the average of the distribution of 345 nm for UV, 550 nm for visible, and 800 nm for infrared (since from Fig. 8.17 of Gates the infrared light distribution transmitted through the atmosphere is shifted notably to shorter wavelengths due to the preferential absorption of the longer wavelengths by clouds). The results for an overcast day for the sea skin layer with and without organic matter are given in table 2.

| OVERCAST SKIES | UV (290-400 nm) | Visible (400-700 nm) | Infrared (700-10000 nm) |
|---|---|---|---|
| Energy flux reaching Earth's surface (Cloud Light, Fig. 3) (total 341.6 W/m$^2$ ) | 22.04 W/m$^2$ (6.5)% | 274.7 W/m$^2$ (80.4%) | 44.86 W/m$^2$ (13.13%) |
| Energy flux absorbed in 1 mm skin layer (pure water). | 0.22x10$^{-3}$ W/ m$^2$ ($A_0^{345} = 1.0x10^{-4}$ cm$^{-1}$) | 0.022 W/m$^2$ ($A_0^{550} = 8.0x10^{-4}$ cm$^{-1}$) | 0.09 W/m$^2$ ($A_0^{800} = 0.02$ cm$^{-1}$) |
| Energy flux absorbed in 1 mm skin layer (ocean water with organic material). | 0.22 W/m$^2$ ($A_{org}^{345} = 0.1$ cm$^{-1}$) | 0.22 W/m$^2$ ($A_{org}^{550} = 8.0x10^{-3}$ cm$^{-1}$) | 0.09 W/m$^2$ ($A_{org}^{800} = 0.02$ cm$^{-1}$) |
| % of total entropy production in skin. | 55.2 | 41.3 | 3.5 |

Table 2. Values of energy deposition per unit time and contribution to the entropy production for the different wavelength regions assuming overcast skies.

Comparing the third row with the second row of table 2 it can be determined that on an overcast day, with the sun directly overhead (air mass of 1.0) the organic matter floating in the surface skin layer increases the absorption of energy in this layer by about 400% over what the value would be without the organic matter, with equal contributions coming from UV and visible light. However, the total energy absorbed in this skin layer on an overcast day is only about 5% that of a clear day, due principally to the comparatively small amount of infrared light that makes it through the clouds. By absorbing and dissipating UV and visible light, the organic matter in this layer contributes about 55.2 + 41.3 = 96.5% to the total entropy production.

By absorbing and dissipating UV and visible light on the surface of oceans and lakes, life therefore augments the entropy production of the Earth in its solar environment. Without life at the surface, a greater portion of light would be reflected, thereby increasing the albedo of Earth (see Clarke et al. (1970) for measurements of the reduction of water albedo due to the presence of organic material), and light would penetrate deeper into the ocean, thereby augmenting the overall blackbody temperature of the ocean (see, for example, Jones et al. (2005) for the effect of phytoplankton on the temperature profile with depth for a lake). Both reduce the entropy production; the former by reducing the amount of available light to dissipate, and the latter by shifting the frequencies of the radiated spectrum to larger values (see Eq. (2)).

Some theories have the origin of life dissipating other sources of free energy, such as chemical energy released from hydrothermal vents at deep ocean trenches. Whether life originated to dissipate the free energy in sunlight or the free energy in made available through chemical transformations, the quantity of life at hydrothermal vents today corresponds to a very minute portion of all life on Earth implying that its contribution to the actual entropy production of the Earth can be considered negligible. The rich ecosystems existing at these vents are, in fact, not completely autonomous, but dependent on the dissolved oxygen and nutrients of photosynthetic life living closer to the surface. An Earth without photosynthetic life would thus correspond to one in a wholly different class of thermodynamic stationary states, one probably with little involvement of a water cycle.

**Evidence for Evolutionary Increases in the Water Cycle**
Plants, far from eliminating transpiration as a wasteful use of free energy, have in fact evolved over time ever more efficient water transport and transpiration systems (Sperry, 2003). There is a general trend in evolution, and in ecosystem succession over shorter times, to ever increasing transpiration rates. For example, conifer forests are more efficient at transpiration than deciduous forests principally because of the greater surface area offered for evaporation by the needles as compared to the leaves. Conifers appeared later in the fossil record (late carbonifourous) and appear in the late successional stage of ecosystems. Root systems are also much more extended in late evolutionary and successional species, allowing them to access water at ever greater depths (Raven and Edwards, 2001). New pigments besides chlorophyll have appeared in the evolutionary history of plants and cyanobacteria, covering an ever greater portion of the intense region of the solar spectrum, even though they have little or no effect on photosynthesis, for example, the carotenoids in plants, or the MAA's found in phytoplankton which absorb across the UVB and UVA regions (310-400 nm) (Whitehead and Hedges, 2002). This is particularly notable in red algae, for example, where its total absorption spectrum has little correspondence with its photosynthetic activation spectrum (Berkaloff *et al.*, 1971).

There exist complex mechanisms in plants to dissipate photons directly into heat, by-passing completely photosynthesis. These mechanisms involve inducing particular electronic de-excitations using dedicated enzymes and proteins and come in a number of distinct classes. Constitutive mechanisms, allow for intersystem crossing of the excited chlorophyll molecule into triplet, long-lived, states which are subsequently quenched by energy transfer to the carotenoids. Inducible mechanisms can be regulated by the plant itself, for example, changing lumen pH causes the production of special enzymes that permit the non-photochemical de-excitation of chlorophyll. Sustained mechanisms are similar to inducible mechanisms but have been adapted to long term environmental stress. For example, over wintering evergreen leaves produce little photosynthesis due to the extreme cold but continue transpiring by absorbing photons and degrading these to heat through non-photochemical de-excitation of chlorophyll. Hitherto, these mechanisms were considered as "safety valves" for photosynthesis, protecting the photosynthetic apparatus against light-induced damage (Niyogi, 2000). However, their existence and evolution can better be understood in a thermodynamic context as augmenting the entropy production potential of the plant through increased transpiration.

The recent findings of microsporine-like amino acids (MAAs) produced by plants and phytoplankton having strong absorption properties in the UVB and UVA regions follows their discovery in fungi (Leach, 1965). They are small (< 400 Da), water-soluble compounds composed of aminocyclohexenone or aminocycloheximine rings with nitrogen or imino alcohol substituents (Carreto et al., 1990) which display strong UV absorption maximum between 310 and 360 nm and high molar extinction (Whitehead and Hedges, 2002). These molecules have been assigned a UV photoprotective role in these organisms, but this appears dubious since more than 20 MAAs have been found in the same organism, each with different but overlapping absorption spectrum, determined by the particular molecular side chain (Whitehead and Hedges, 2002). If their principle function were photoprotective, there existence would be confined to those UV wavelengths that cause damage to the organism, and not to the whole UV broadband spectrum.

Plants also perform a free energy intensive process known as photorespiration in which $O_2$ instead of $CO_2$ is captured by the binding enzyme RuBisCo, the main enzyme of the light-independent part of photosynthesis. This capture of $O_2$ instead of $CO_2$ (occurring about 25% of the time) is detrimental to the plant for a number of reasons, including the production of toxins that must be removed (Govindjee, 2005) and does not lead to ATP production. There is no apparent utility to the plant in performing photorespiration and in fact it reduces the efficiency of photosynthesis. It has often been considered as an "evolutionary relic" (Niyogi, 2000), still existing from the days when $O_2$ was less prevalent in the atmosphere than today and $CO_2$ more so (0.78% $CO_2$ by volume at the rise of land plants during the Ordovician (ca. 470 Ma) compared with only 0.038% today). However, such an explanation is not in accord with the known efficacy of natural selection to eliminate useless or wasteful processes. Another theory has photorespiration as a way to dissipate excess photons and electrons and thus protect the plants photosynthesizing system from excess light-induced damage (Niyogi, 2000). Since photorespiration is common to all $C_3$ plants, independent of their insolation environments, it is more plausible that photorespiration, being completely analogous to photosynthesis with respect to the dissipation of light into heat in the presence of water (by quenching of excited chlorophylls) and subsequent transpiration of water, is retained for its complimentary role in evapotranspiration and thus entropy production.

Plants not only evaporate water during sunlight hours, but also at night (Snyder et al., 2003). Common house plants evaporate up to 1/3 of the daily transpired water at night (Hernández Candía, 2009). Not all the stoma in $C_3$ and $C_4$ photosynthetic plants are closed at night and some water vapor also diffuses through the cuticle at night. The physiological reason, in benefit of the plant, for night transpiration, if one exists, remains unclear. It, of course, can have no relevance to cooling leaves for optimal photosynthetic rates. Explanations range from improving nutrient acquisition, recovery of water conductance from stressful daytime xylem cavitation events, and preventing excess leaf turgor when water potentials become large during the day (Snyder et al., 2003). However, night transpiration is less of an enigma if considered as a complement to the thermodynamic function of life to augment the entropy production of Earth through the water cycle. In this context, it is also relevant that chlorophyll has an anomalous absorption peak in the infrared at between about 4,000 and 10,000 nm (Gates, 1980), close to the wavelength at which the blackbody radiation of the Earth's surface at 14 °C peaks.

Cyanobacteria have been found to be living within Antarctic ice at depths of up to 2 m. These bacteria and other lithogenic material absorb solar radiation which causes the formation of liquid water within the ice even though the outside air temperatures may be well below freezing. This heating from below causes excess ablation and sublimation of the overlying ice at rates as high as 30 cm per year (Priscu *et al.*, 2005).

Finally, by analyzing latent heat fluxes (evaporation) and the $CO_2$ flux for plants from various published data sets, Wang et al. (2007) have found vanishing derivatives of transpiration rates with respect to leaf temperature and $CO_2$ flux, suggesting a maximum transpiration rate for plants, i.e. that the particular partition of latent and sensible heat fluxes is such that it leads to a leaf temperature and leaf water potential giving maximal transpiration rates, and thus maximal production of entropy (Wang et al., 2007).

**The Function of Animals**
If the primary thermodynamic function of the plants and cyanobacteria is to augment the entropy production of the Earth by absorbing light in the presence of liquid water, it may then be asked: What is the function of higher mobile animal life?

Because of their intricate root system which allows the plants to draw up water for evaporation from great depths, plants are not mobile and depend on insects and other animals for their supply of nutrients, cross fertilization, and seed dissemination and dispersal into new environments. The mobility and the short life span of insects and animals mean that through excrement and eventual death, they provide a reliable mechanism for dispersal of nutrients and seeds.

Crustaceans and animal marine life in water perform a similar function as insect and animal life on land. These higher forms of life distribute nutrients throughout the ocean surface through excrement and dying. It is noteworthy that dead fish and mammals do not sink rapidly to the bottom of the sea, but remain floating for considerable time on the surface where, as on land, bacteria break down the organism into its components, allowing photon dissipating phytoplankton to reuse the nutrients, particularly nitrogen. It is interesting that many algae blooms produce a neurotoxin with apparently no other end than to kill higher marine life. There is also a continual cycling of nutrients from the depths of the ocean to the surface as deep diving mammals preying on bottom feeders release nutrients at the surface through excrement and death. Because of this cycling and mobility of animals, a much larger portion of the ocean surface is rendered suitable for phytoplankton growth, offering a much larger area for efficient surface absorption of sunlight and evaporation of water than would otherwise be the case.

From this thermodynamic viewpoint, animal life provides a specialized gardening service to the plants and cyanobacteria, which in turn catalyze the absorption and dissipation of sunlight in the presence of water, promoting entropy production through the water cycle. There is strong empirical evidence suggesting that ecosystem complexity, in terms of species diversity, is correlated with potential evapotranspiration (Gaston, 2000). The traditional ecological pyramid should thus be turned on its pinnacle. Instead of plants and phytoplankton being considered as the base that sustains animal life, animals are in fact the unwitting but content servants of plant and phytoplankton life, obtaining thermodynamic relevance only in how they increase the plant and phytoplankton potential for evaporation of water.

**Conclusions**
We have argued that the basic thermodynamic function of life (and organic material in general) is to absorb and dissipate high energy photons such that the heat can be absorbed by liquid water and eventually transferred to space through the water cycle. Photosynthesis, although relevant to cyanobacteria and plant growth, has only minor relevance to the thermodynamic function of life. Augmenting the water cycle through increased photon absorption and radiationless relaxation, life augments the entropy production of the Earth in its interaction with its solar environment. We have presented empirical evidence indicating that the evolutionary history of Earth's biosphere is one of increased photon absorption and dissipation over time, whether on shorter successional, or longer evolutionary, time scales.

This thermodynamic perspective on life views it as a catalyst for entropy production through the water cycle, and ocean and wind currents. It ties biotic processes to abiotic processes with the universal goal of increasing Earth's global entropy production and thus provides a framework within which coevolution of the biotic with the abiotic can be accommodated. In important distinction to the hypothesis of Gaia, that mixed biotic-abiotic mechanisms have evolved to maintain the conditions on Earth suitable to life, it is here suggested instead that these biotic-abiotic mechanisms have evolved to augment the entropy production of Earth, principally, but not exclusively, through the facilitation of the water cycle. Life, as we know it, is an important, perhaps even inevitable, but certainly not indispensable, catalyst for the production of entropy on Earth.

**Acknowledgements**

The author is grateful for the financial assistance of DGAPA-UNAM, grant numbers IN118206 and IN112809.